\documentclass[12pt]{article}
\usepackage[top=1.15in, bottom=1.4in, left=1in, right=1in]{geometry}

\usepackage[utf8]{inputenc}
\usepackage{enumerate,setspace,graphicx,epstopdf,amsmath,amsfonts,amssymb,amsthm}
\usepackage{marginnote,datetime,enumitem,subfigure,rotating,fancyvrb}

\usepackage[hang]{footmisc}
\usepackage{natbib}
\usepackage{authblk}
\usepackage{threeparttable}
\usepackage{makecell}
\usepackage[dvipsnames]{xcolor}
\usepackage{rotating}

\usepackage{array}
\newcolumntype{L}[1]{>{\raggedright\let\newline\\\arraybackslash\hspace{0pt}}m{#1}}
\newcolumntype{C}[1]{>{\centering\let\newline\\\arraybackslash\hspace{0pt}}m{#1}}
\newcolumntype{R}[1]{>{\raggedleft\let\newline\\\arraybackslash\hspace{0pt}}m{#1}}

\usepackage[hang]{footmisc}
\usepackage{natbib}
\usepackage{authblk}
\usepackage{threeparttable}
\usepackage{makecell}
\usepackage[dvipsnames]{xcolor}

\usepackage{array}
\usepackage{tabularx}
\newcolumntype{Y}{>{\centering\arraybackslash}X}

\usepackage{indentfirst}

\usepackage{epsfig}

\usepackage{indentfirst}

\usepackage{epsfig}

\begin{document}

\title{\vspace{-2.5cm} \textbf{Social Distancing Beliefs and Human Mobility: Evidence from Twitter }}
\date{\today}
	\author[a]{Simon Porcher\thanks{Electronic address: \texttt{simon.porcher@univ-paris1.fr}; IAE Paris - Université Paris 1 Panth\'eon-Sorbonne, 8 bis rue Croix de Jarry, 75013 Paris.}}
	\author[b]{Thomas Renault\thanks{Email: \texttt{thomas.renault@univ-paris1.fr}; Corresponding author: Thomas Renault. CES Sorbonne - Université Paris 1 Panth\'eon-Sorbonne, Maison des Sciences Économiques. 106-112, boulevard de l'Hôpital 75013 Paris.}}

\affil[a]{{ \small IAE Paris, Universit\'e Paris 1 Panth\'eon Sorbonne} \vspace{-0.1in}}
\affil[b]{{ \small Universit\'e Paris 1 Panth\'eon Sorbonne, CES}
\vspace{-0.1in}}

	\maketitle

\vspace{-0.25in}

\begin{abstract}
\noindent

We construct a novel database containing hundreds of thousands geotagged messages related to the COVID-19 pandemic sent on Twitter. We create a daily index of social distancing -- at the state level -- to capture social distancing beliefs by analyzing the number of tweets containing keywords such as "stay home", "stay safe", "wear mask", "wash hands" and "social distancing". We find that an increase in the Twitter index of social distancing on day t-1 is associated with a decrease in mobility on day t. We also find that state orders, an increase in the number of COVID cases, precipitation and temperature contribute to reducing human mobility. Republican states are also less likely to enforce social distancing. Beliefs shared on social networks could both reveal the behavior of individuals and influence the behavior of others. Our findings suggest that policy makers can use geotagged Twitter data -- in conjunction with mobility data -- to better understand individual voluntary social distancing actions.

\bigskip
\noindent \textbf{Keywords}: COVID-19, Social Distancing, Beliefs, Human Mobility, Twitter. \\\\
\noindent \textbf{JEL classification}:D1-D83-I18-L82.
\end{abstract}

\thispagestyle{empty}
\clearpage

\doublespacing

\section{Introduction}

Social distancing policies reduce social interactions and ultimately COVID-19 infections. Epidemiologists such as \cite{ferguson2020} estimate that the implementation of social distancing -- including case isolation, household quarantine and school and workplace closures -- could halve the number of deaths in the United Kingdom and the United States. A growing body of literature has linked policy interventions with social distancing \citep{abouk, gupta2020, chernozhukov2020} and the latter with the spread of contamination \citep{kapoor2020god, chernozhukov2020, yilmazkuday2020stay}. While evidence shows that government interventions decrease the size of the pandemic and redistribute the number of cases over time, little empirical research has explored the impact of beliefs on social distancing \citep{allcott2020polarization}.

In this paper, we contribute to the emerging literature studying differences in social distancing across the United States during the COVID-19 pandemic. We proxy for the beliefs of agents by creating a Twitter index of social distancing based on geotagged tweets posted between February and June 2020. Twitter-based measures \citep{mooreetal} -- like newspaper-based measures \citep{altigetal} or Google search trends \citep{brodeur2020covid} -- are considered good proxies for the perceptions and feelings of households. Moreover, collecting tweets avoids the small-sample biases that can be found in most studies based on questionnaires. Previous work using social network data shows that they successfully predict some economic outcomes \citep{renault2017}, disease outbreaks \citep{carneiromylonakis} or happiness \citep{brodeur2020covid}.

We relate our Twitter index of social distancing to measures of mobility computed by Google at the state level for 49 US States\footnote{We remove Alaska and Hawaii from our analysis, and we include Washington DC.}, controlling for the dates of implementation of the various state orders (stay-at-home orders, school closures and nonessential business closures). We find strong evidence that differences in the Twitter index of social distancing correlate with differences in mobility between states, even after controlling for the various dates of implementations of state orders, rainfall, temperature and the number of new COVID cases. Our results show that a substantial voluntary response of agents cannot be explained by government responses to the COVID-19 outbreak.

The results of the paper are of interest for researchers working on spatial differentiation in the human response to social distancing. A range of papers have studied the relationship between social capital and social distancing in the United States. \cite{dingetal} show that measures of social capital, such as community engagement, moderate the effect of statewide mobility restrictions on social distancing. For example, community engagement implies greater costs of social distancing and decreases the impact of stay-at-home orders on social distancing. \cite{barriosetal} use civic capital as a moderator of stay-at-home orders to explain compelled social distancing in the US, at both the individual and country level, and in Europe. They find that a higher sense of civic duty leads to greater compliance with social distancing rules, even after the end of a domestic lockdown.

More related to our paper, \cite{allcott2020polarization} model differences in beliefs and attitudes as resulting from messages on the crisis from both political leaders and the media. They use survey data and show that Republicans and Democrats engage in social distancing to different extents. Their theoretical model is also validated by \cite{bursztynetal} and \cite{simonovetal}, who study the impact of Fox News on stay-at-home behaviors and show that greater exposure leads to less compliance with social distancing. The present paper contributes to the literature on the impact of media on compliance with social distancing rules, as social media not only signals the behavior or sentiment of the population but also has an impact on readers.

The paper is also related to a growing body of literature on the impact of COVID-19 on household uncertainty. \cite{altigetal} document a huge increase in uncertainty before and after the COVID-19 pandemic, using various indicators of economic uncertainty, including newspaper scraping and measures from expectations surveys. In the same vein, \cite{bakertwitter} construct a Twitter-based economic uncertainty index scraping worldwide tweets containing the keywords 'economic' and 'uncertainty' in the first semester of 2020 to obtain alternative measures of economic policy uncertainty to measure volatility. Our Twitter-based measure of social distancing captures concerns by the population at the state level, and the methodology we use can be replicated by researchers seeking to study the effect of information on behaviors during COVID-19 in different countries.\footnote{Our Twitter-based measure of social distancing is available online: http://www.xxxxxxxxx.com}

Finally, the paper is of interest for researchers working on well-being in the era of COVID-19. \cite{brodeur2020covid} study the impact of lockdown policies on happiness, as measured by keyword searches in Google Trends, in Europe and America. They find that people's mental health may have been severely affected by the lockdown. The paper that is the closest to ours is that by \cite{alfaroetal}. They study mobility in 89 cities worldwide as an outcome of fear measured by Google search data, some measures of social preferences and government lockdowns. They find that both lockdown policies and fear have a negative effect on mobility. Our paper contributes to this stream of the literature by using Twitter-based scraped data, rather than Google Trends data, which give only relative numbers rather than absolute numbers. The advantage of scraping is that it allows us to obtain the absolute numbers and to select the most appropriate tweets related to social distancing. Google Trends does not allow us to create an alternative Google index sorting out search terms that are correlated with a given term but not having the same meaning (e.g., fear and the TV show ``Fear Factor'').

The remainder of the paper is organized as follows. Sections \ref{data} and \ref{method} present the data and the method, respectively. Section \ref{result} introduces the results, and Section \ref{conclusion} concludes the paper.


\section{Data}
	\label{data}
\subsection{Mobility}

We use daily Community Mobility Reports data from Google as a proxy for social distancing. The reports chart movement trends over time by geography, across different categories of places such as retail and recreation, groceries and pharmacies, parks, transit stations, workplaces, and residential areas. The mobile location data -- anonymized and aggregated at the state level in the United States -- are available since Feb 15, 2020, and show how visitors to categorized places change compared to the baseline period of Jan 3 to Feb 6, 2020 (before the COVID-19 outbreak). The residential category shows a change in time spent at home. The five other categories measure the change in total visitors to categorized places. The data exhibit large differences over time and space, both in levels and in variations.

\subsection{Twitter Data}

We use geotagged messages from Twitter to capture beliefs at the state level.\footnote{We focus on the state level instead of the county level, as the number of tweets we collect was too low to construct a reliable daily indicator of information at the county level} To construct our database, we use a web scraping tool, and we extract all tweets containing the following keywords: "stay home", "stay safe", "save lives", "wash hands", "wear mask" and "social distancing"\footnote{We also consider variants of those terms and hashtags such as \#StayHome}. We find that the number of messages protesting against social distancing measures is very low: the percentage of tweets containing keywords or hashtags such as \#ReOpenAmerica, \#LockdownProtest (or related keywords) represents less than 1\% of all tweets in our sample. As more than 99\% of the tweets containing social distancing keywords are encouraging social distancing, we do not use a sentiment analysis algorithm to derive the polarity of each message.

We focus our attention on geotagged tweets -- messages for which the location of the user is known -- to construct a daily indicator of social distancing beliefs at the state level. We define $TwitterSD_{s,t}$ as the number of tweets about social distancing sent by users located in state "s" on day t for 100,000 inhabitants:

\begin{equation}
	TwitterSD_{s,t} = \dfrac{Number of Social Distancing Tweet_ {s,t}}{Population_s} * 100,000
	\end{equation}

\begin{figure}[]
\begin{center}
\caption[]{Example of a tweet encouraging social distancing.}
\vspace{5mm}
\begin{center}
\epsfig{file=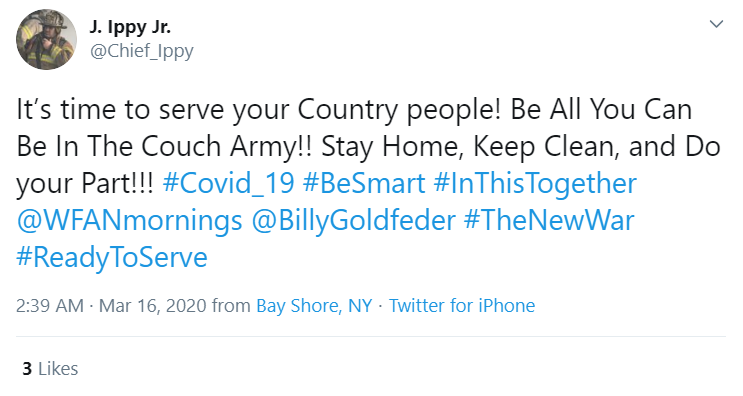, width=0.6 \textwidth}
\end{center}
\vspace*{-2mm}
{\scriptsize Notes: This figure presents an example of a tweet encouraging social distancing. This message reveals the beliefs of the user who sends it and could also influence the behaviors of other users reading the tweet. This tweet was sent on March 16, 2020, by a user located in Bay Shore, New York. This tweet received three likes. We use the number of likes as a proxy for influence when we construct our weighted Twitter-based index.}
\vspace*{10mm}
			\label{tweet-example}
\end{center}
\end{figure}

We also construct a weighted Twitter-based index by weighting each message encouraging social distancing by the logarithm of 1 plus its number of likes, its number of retweets or its number of replies. Tweets with higher numbers of likes, retweets and replies could have a greater impact on the reduction of mobility, as these messages will be disseminated more widely and may impact the beliefs of other users. Figure \ref{tweet-example} presents an example of a tweet in our sample where a user both reveals his belief and encourages other users to respect social distancing.

Twitter is widely used across the United States: there is a total of nearly 50 million monthly active Twitter users in the US, and 20 million Twitter users are on the platform daily. Furthermore, users on Twitter mostly follow users from the same metropolitan area. According to \cite{takhteyev2012geography}, 39 percent of ties connect users within the same regional cluster. This suggests that users in a given state are more exposed to the tweets and beliefs of other users located in the same state. One of the main drawbacks of using data from Twitter is that geotagged tweets are not representative of the US population: the Twitter population is biased towards higher incomes and urban areas \cite{malik2015population}, and geotagged tweets are written more often by young people and by women \cite{pavalanathan2015confounds}. While we acknowledge that this could limit the generalizability of our findings, we add state fixed effects, and we focus on variation across states and over time to limit the bias due to the specific sample of Twitter users.

\begin{figure}[!ht]
\centering
\caption[]{[Appendix A] Twitter - Number of Social Distancing tweets per day.}
\vspace{5mm}
\begin{center}
\epsfig{file=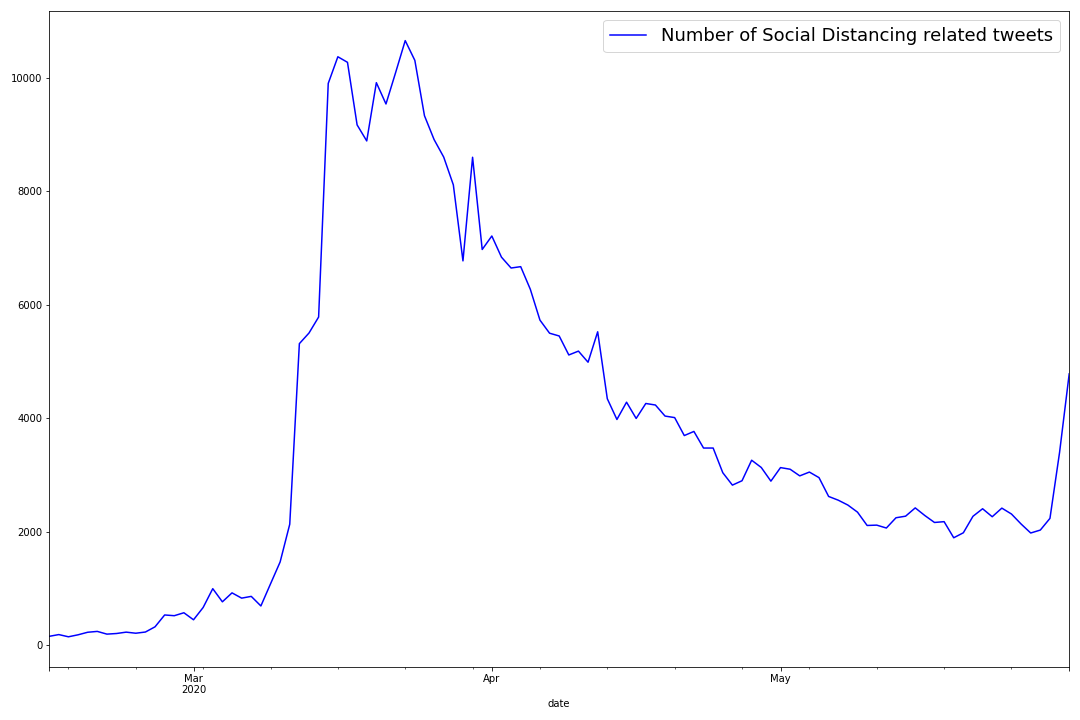, width=1 \textwidth}
\end{center}
\vspace*{-2mm}
{\scriptsize Notes: This Figure presents the number of tweets per day encouraging social distancing behaviors.}
\vspace*{10mm}
\label{AppendixA}
\end{figure}

Geotagged messages only represent 1 to 2\% of all messages sent on Twitter every day. However, given the very large number of messages sent every day on the platform (approximately 500 million tweets), we still obtain a large database of 402,005 messages containing at least one social distancing keyword sent between February 15 and May 31, 2020. Tweets in our sample have an average of 9.21 likes, 1.84 retweets and 0.62 replies. Appendix \ref{AppendixA} shows the evolution of the total number of tweets related to social distancing during our sample period. We observe that the number of tweets increases sharply during the second week of March -- a week, coinciding with the declaration by the World Health Organization stating that the COVID-19 outbreak was effectively a pandemic and with a structural break in mobility series identified by \cite{cronin2020private}.

\subsection{Controls}

We used epidemiological data from the Center for Systems Science and Engineering (CSSE) at Johns Hopkins University. We compute the number of new COVID cases per 100,000 inhabitants for each state and day.\footnote{We also consider the number of deaths per 100,000 inhabitants, and we find similar results}. We also consider state-level social distancing policies from \cite{fullman2020state}. The dataset includes a wide set of policies, such as restrictions on gatherings, school closures, stay-at-home orders, and nonessential business closures. We create a dummy variable by state to indicate whether any of the previous policies were in place on a given date. We consider the date of policy enactment. We also create a variable to capture political polarization by considering the percentage of Trump votes by state during the 2016 election. Finally, we use environmental data from the National Centers for Environmental Information to control for daily rainfall and temperature in each state. We use the average level of rainfall and the average maximum temperature by considering observations from all weather stations located in each state.

\section{Methods}
	\label{method}

We use a simple causal framework in which individuals make their behavioral decisions based on the marginal benefits and costs of interaction \citep{andersen}, which depends on their beliefs and the information they have. The COVID-19 outbreak increases the marginal costs of social interaction by increasing the probability of infection. This probability of infection is localized and differs across geographic areas. We estimate the following OLS model that summarizes our conceptual framework:

\begin{equation}
	Mobility_{s,t} = \beta_1 TwitterSD_{s,t-1} + X_{s,t} + \delta_{s} + \delta_{d,t} + \epsilon_{s,t}
	\label{model1}
	\end{equation}

where $\beta_1$ captures the effect of our Twitter social distancing (SD) index.\footnote{The results are similar when we consider $TwitterSD_{s,t}$ instead of $TwitterSD_{s,t}$. We choose to use the lagged value of our Twitter index in an effort to limit the reverse causality between time spent at home and the number of tweets sent (as users might tweet more when they have more spare time).} Different communities have different perceptions of risks \citep{andersen} and may use social media to reveal their beliefs and influence the behavior of other agents. $X_{s,t}$ is a vector of state-level time-varying controls including average temperature and rainfall, the number of new COVID cases and state-level social distancing policies. External factors, such as weather conditions, affect the marginal benefits and costs of social interaction. By the same token, government policies implemented to increase the costs of social interaction are important drivers of social distancing. $\delta_{s}$ controls for all time-invariant state characteristics, such as population density, preferences or income. For example, different communities have different preferences in terms of social interactions and risky behaviors. Thus, the marginal cost of social distancing depends on how much people value outside gatherings, traveling or working from home, and on their private risk of infection, e.g., whether they suffer from chronic diseases or the number of COVID-19 cases in the community. $\delta_{d,t}$ controls for time-varying division characteristics as well as all time-varying national and international factors. We consider the nine divisions from the United States Census Bureau.\footnote{We also consider the 4 US regions (Northeast, Midwest, South, West), and we find similar results}

\section{Results}
	\label{result}

Our baseline results are reported in Table \ref{Table1}. The first column reports the OLS model for residential mobility, and columns (2) to (6) show the results for mobility in various places. All models include stated fixed effects and division$\times$time fixed effects. We find that the coefficient of our Twitter index of social distancing is positive in column (1) and negative in columns (2) to (6). One additional tweet encouraging social distancing per 100,000 inhabitants -- which represents an increase of approximately 0.66 standard deviations -- is associated with an increase in time spent at home of approximately 0.3 \%. The absolute size of the coefficient is on average lower for residential mobility, as the residential category shows a change in duration, while the other categories measure a change in total visitors.\footnote{As people already spend a large portion of their time at home (even outside the COVID period), the capacity for variation is limited.} We also find that the magnitude of the effect is approximately three times larger for mobility to workplaces, transit stations and national parks than for mobility to retail areas and grocery stores. All of the other variables are of the expected sign: the number of COVID cases that proxies for the marginal infection probability decreases all types of mobility (or increases the time spent at home), rainfall increases time spent home and decreases mobility to parks and grocery stores, and temperatures decrease time spent home and increase mobility to other places. As expected, stay-at-home orders have a large and significant effect on mobility. Other social distancing measures, such as school closures, nonessential business closures, and gathering restrictions, do not significantly impact mobility when we include stay-at-home orders and division*time fixed effects. Although these controls attenuate the effect of the Twitter index of social distancing to some degree, the latter remains significant in all specifications.\\

Table \ref{Table2} presents our results when we weight tweets by the number of likes, the number retweets or the number of replies. Column (1) reports the results of the baseline model. Columns (2), (3) and (4) report the results of the model using the Twitter index of social distancing weighted by the number of likes, retweets and replies, respectively. The Twitter indices are standardized to ensure comparability across the different specifications. The results are only presented for time spent at home to illustrate the impact of the weighted Twitter indices. The results for the other types of mobility are reported in Appendix \ref{AppendixB}. We find that the effects are relatively similar when we use weighted tweets instead of unweighted tweets. The relative sign of the coefficients of our baseline measure and the weighted measure suggest that the relation between the number of messages related to social distancing and the observed reduction in mobility is mostly driven by self-disclosed beliefs rather than by the influence of local tweets on other users’ behavior. This result might also be driven by the fact that social media tends to favor online ``bubbles'': users on Twitter are exposed primarily to ideologically similar information \citep{eady2019many}, and thus, the influence of tweets encouraging social distancing sent by individual users might be limited. \\

An important control for beliefs is partisanship. Table \ref{Table3} presents our results when we add an interaction between the share of Republican votes in the 2016 US presidential election and stay-at-home orders. We find -- as in \cite{allcott2020polarization} -- that states with a larger share of Republican votes in 2016 tend to enforce less social distancing. The interaction is significant in all the specifications and of the opposite sign as stay-at-home orders. A greater share of republican votes decreases the impact of stay-at-home orders on mobility. This validates the intuition that the perception of risk differs significantly based on the partisanship of the community. Intuitively, Republicans might be more attached to individual freedom than to state action, while Democrats might overreact to state orders.\footnote{We also interact our Twitter-based indicator with the share of Republican votes in the 2016 US presidential election. The interaction is not significant, suggesting that the effect is similar for both Democrats and Republicans.} The coefficient on the Twitter index of social distancing remains stable and is even slightly higher when partisanship is added. \\

Table \ref{Table4} presents the same model as in Table \ref{Table3} with various geographic and time fixed effects. The sign and magnitude of the coefficients of social distancing tweets on the time spent at home are similar to those in Tables \ref{Table2} and \ref{Table3}, ranging between 0.351 and 1.134. The models including only geographic fixed effects yield upward-biased coefficients, while the models including both time and regional fixed effects yield coefficients that are more representative of the real impact of beliefs on mobility, as individuals tend to adapt their behavior in terms of mobility but also on Twitter over time. \\

\begin{sidewaystable}
\centering
\caption{Baseline Model - Impact of the Twitter index of social distancing on mobility}
\vspace{0.5cm}
\begin{tabular}{L{5cm}C{2.5cm}C{2.5cm}C{2.5cm}C{2.5cm}C{2.5cm}C{2.5cm}}
\hline
                               & {[}1{]}     & {[}2{]}    & {[}3{]}          & {[}4{]}    & {[}5{]}             & {[}6{]}              \\
 VARIABLES & Residential & Workplaces & Grocery and pharmacy & Retail and recreation & Transit stations & Parks \\ \hline
 &  &  &  &  &  &  \\
Social Distancing Tweets & 0.295*** & -0.827*** & -0.235* & -0.319** & -1.047*** & -1.141*** \\
 & (0.0570) & (0.115) & (0.129) & (0.152) & (0.201) & (0.378) \\
New COVID cases & 0.0927*** & -0.150*** & -0.118** & -0.227*** & -0.256*** & -0.598*** \\
 & (0.0204) & (0.0407) & (0.0551) & (0.0726) & (0.0935) & (0.216) \\
Stay-at-home order & 1.012*** & -2.075*** & -3.179*** & -2.759** & -6.864*** & -14.13** \\
 & (0.350) & (0.710) & (0.867) & (1.136) & (2.079) & (5.806) \\
School Closure & 0.232 & -0.306 & -0.560 & -1.912 & 0.00216 & -2.684 \\
 & (0.270) & (0.654) & (1.030) & (1.315) & (0.971) & (4.335) \\
Gathering restrictions & 0.251 & -1.244*** & 0.695 & -1.043 & 0.917 & 7.423 \\
 & (0.229) & (0.455) & (0.755) & (1.002) & (1.853) & (6.092) \\
Business closure & 0.385 & -0.924 & -0.644 & -1.779 & 0.676 & 6.328 \\
 & (0.569) & (1.048) & (1.380) & (1.577) & (2.480) & (6.933) \\
Precipitation & 0.00290*** & -0.00191 & -0.00313* & -0.00312 & -0.00329 & -0.0534*** \\
 & (0.000658) & (0.00129) & (0.00160) & (0.00202) & (0.00277) & (0.0123) \\
Temperature & -0.0118*** & 0.00612** & 0.0367*** & 0.0267*** & 0.0277** & 0.316*** \\
 & (0.00144) & (0.00294) & (0.00433) & (0.00515) & (0.0110) & (0.0300) \\
Constant & -1.061*** & -0.501 & -2.850 & 7.895** & 8.187*** & 14.04 \\
 & (0.389) & (1.292) & (3.436) & (3.328) & (1.981) & (13.65) \\
 &  &  &  &  &  &  \\
State FE & Yes & Yes & Yes & Yes & Yes & Yes \\
 Division*Time FE & Yes & Yes & Yes & Yes & Yes & Yes \\
 Observations & 5,194 & 5,194 & 5,194 & 5,194 & 5,194 & 5,183 \\
R-squared & 0.975 & 0.981 & 0.920 & 0.967 & 0.945 & 0.824 \\\hline
\hline
\end{tabular}
\begin{minipage}[!hb]{1\linewidth}
\vspace{0.5cm}
Note: All models are OLS regressions with state fixed effects and division*time fixed effects. Model (1) uses the time spent at home from Google Mobility. Models (2) to (5) use Google Mobility data for various venues. State-level clustered robust standard errors in parentheses with *** p$<$0.01, ** p$<$0.05, * p$<$0.1. 
\end{minipage}
\label{Table1}
\end{sidewaystable}

\begin{sidewaystable}
\centering
\caption{Tweet weighted model - Baseline model compared with tweets weighted by likes, retweets and replies}
\vspace{0.5cm}
\begin{tabular}{L{5cm}C{2.5cm}C{2.5cm}C{2.5cm}C{2.5cm}}
\hline
& (1) & (2) & (3) & (4) \\
VARIABLES & Residential & Residential & Residential & Residential \\ \hline
 &  &  &  &  \\
SD Tweets & 0.435*** &  &  &  \\
 & (0.0841) &  &  &  \\
SD Tweets * likes &  & 0.292*** &  &  \\
 &  & (0.0532) &  &  \\
SD Tweets * retweets &  &  & 0.292*** &  \\
 &  &  & (0.0403) &  \\
SD Tweets * replies &  &  &  & 0.281*** \\
 &  &  &  & (0.0425) \\
New COVID cases & 0.0927*** & 0.0944*** & 0.0938*** & 0.0941*** \\
 & (0.0204) & (0.0205) & (0.0202) & (0.0202) \\
Stay-at-home order & 1.012*** & 1.004*** & 1.011*** & 0.992*** \\
 & (0.350) & (0.347) & (0.345) & (0.344) \\
School Closure & 0.232 & 0.234 & 0.203 & 0.250 \\
 & (0.270) & (0.272) & (0.270) & (0.273) \\
Gathering restrictions & 0.251 & 0.270 & 0.274 & 0.277 \\
 & (0.229) & (0.227) & (0.229) & (0.225) \\
Business closure & 0.385 & 0.403 & 0.400 & 0.407 \\
 & (0.569) & (0.577) & (0.574) & (0.574) \\
Precipitation & 0.00290*** & 0.00293*** & 0.00295*** & 0.00294*** \\
 & (0.000658) & (0.000645) & (0.000657) & (0.000630) \\
Temperature & -0.0118*** & -0.0120*** & -0.0120*** & -0.0120*** \\
 & (0.00144) & (0.00145) & (0.00146) & (0.00146) \\
Constant & -0.753* & -0.873** & -0.894** & -0.881** \\
 & (0.393) & (0.399) & (0.399) & (0.403) \\
 &  &  &  &  \\
State FE & Yes & Yes & Yes & Yes \\
 Division*Time FE & Yes & Yes & Yes & Yes \\
 Observations & 5,194 & 5,194 & 5,194 & 5,194 \\
R-squared & 0.975 & 0.975 & 0.975 & 0.975 \\ \hline             
\end{tabular}
\begin{minipage}[!hb]{1\linewidth}
\vspace{0.5cm}
Note: All models are OLS regressions with state fixed effects and division*time fixed effect and use the time spent at home from Google Mobility. The Twitter indices of social distancing - $Social Distancing Tweets$, $Social Distancing Tweets\_likes$, $Social Distancing Tweets\_retweets$, $Social Distancing Tweets\_replies$ - are standardized. State-level clustered robust standard errors in parentheses with *** p$<$0.01, ** p$<$0.05, * p$<$0.1. 
\end{minipage}
\label{Table2}
\end{sidewaystable}

\begin{sidewaystable}
\centering
\caption{Baseline model with an interaction between stay-home orders and Republicans votes}
\vspace{0.5cm}
\begin{tabular}{L{5cm}C{2.5cm}C{2.5cm}C{2.5cm}C{2.5cm}C{2.5cm}C{2.5cm}} \hline
                              & {[}1{]}     & {[}2{]}    & {[}3{]}          & {[}4{]}    & {[}5{]}             & {[}6{]}              \\
 VARIABLES & Residential & Workplaces & Grocery and pharmacy & Retail and recreation & Transit stations & Parks \\ \hline
Social Distancing Tweets & 0.351*** & -0.930*** & -0.348** & -0.466** & -1.239*** & -1.502*** \\
 & (0.0444) & (0.0893) & (0.159) & (0.215) & (0.176) & (0.382) \\
New COVID cases & 0.0572*** & -0.0849*** & -0.0468 & -0.134*** & -0.136** & -0.371** \\
 & (0.0133) & (0.0226) & (0.0356) & (0.0430) & (0.0650) & (0.180) \\
Stay-at-home order & 6.940*** & -12.95*** & -15.01*** & -18.12*** & -26.99*** & -51.84*** \\
 & (0.780) & (1.493) & (1.872) & (2.655) & (4.322) & (10.95) \\
Stay-at-home order * Rep & -11.93*** & 21.89*** & 23.81*** & 30.90*** & 40.49*** & 75.93*** \\
 & (1.237) & (2.412) & (3.049) & (5.165) & (7.333) & (16.49) \\
School Closure & 0.122 & -0.106 & -0.341 & -1.629 & 0.373 & -1.991 \\
 & (0.283) & (0.700) & (1.142) & (1.509) & (1.212) & (4.742) \\
Gathering restrictions & 0.390* & -1.500*** & 0.417 & -1.404 & 0.444 & 6.532 \\
 & (0.200) & (0.427) & (0.786) & (0.985) & (1.852) & (6.129) \\
Business closure & 0.239 & -0.655 & -0.351 & -1.399 & 1.175 & 7.243 \\
 & (0.409) & (0.754) & (1.115) & (1.164) & (2.046) & (6.434) \\
Precipitation & 0.00267*** & -0.00150 & -0.00268 & -0.00254 & -0.00253 & -0.0519*** \\
 & (0.000726) & (0.00140) & (0.00170) & (0.00210) & (0.00304) & (0.0125) \\
Temperature & -0.0116*** & 0.00580** & 0.0364*** & 0.0263*** & 0.0271** & 0.315*** \\
 & (0.00148) & (0.00281) & (0.00430) & (0.00521) & (0.0110) & (0.0306) \\
Constant & -1.968*** & 1.164 & -1.039 & 10.25*** & 11.27*** & 19.82 \\
 & (0.242) & (0.985) & (3.137) & (2.874) & (1.559) & (13.02) \\
 &  &  &  &  &  &  \\
 State FE & Yes & Yes & Yes & Yes & Yes & Yes \\
 Division*Time FE & Yes & Yes & Yes & Yes & Yes & Yes \\ 
Observations & 5,194 & 5,194 & 5,194 & 5,194 & 5,194 & 5,183 \\
R-squared & 0.979 & 0.983 & 0.926 & 0.971 & 0.950 & 0.828 \\
\hline
\end{tabular}
\begin{minipage}[!hb]{1\linewidth}
\vspace{0.5cm}
Note: All models are OLS regressions with state fixed effects and division*time fixed effects. Model (1) uses the time spent at home from Google Mobility. Models (2) to (5) use Google Mobility data for various venues. State-level clustered robust standard errors in parentheses with *** p$<$0.01, ** p$<$0.05, * p$<$0.1. 
\end{minipage}
\label{Table3}
\end{sidewaystable}

\begin{sidewaystable}
\centering
\caption{Regression with different combinations set of fixed-effects}
\vspace{0.5cm}
\begin{tabular}{L{5cm}C{2.5cm}C{2.5cm}C{2.5cm}C{2.5cm}C{2.5cm}C{2.5cm}}
\hline
 & (1) & (2) & (3) & (4) & (5) & (6) \\
VARIABLES & Residential & Residential & Residential & Residential & Residential & Residential \\ \hline
 &  &  &  &  &  &  \\
Social Distancing Tweets & 1.029** & 0.499*** & 1.134** & 0.361*** & 0.351*** & 0.374*** \\
 & (0.411) & (0.124) & (0.443) & (0.0522) & (0.0444) & (0.0412) \\
New COVID cases & 0.0906*** & 0.122*** & 0.0433* & 0.0658*** & 0.0572*** & 0.0642*** \\
 & (0.0311) & (0.0256) & (0.0247) & (0.0147) & (0.0133) & (0.0159) \\
Stay-at-home order & 8.808*** & 9.419*** & 9.514*** & 7.121*** & 6.940*** & 7.326*** \\
 & (1.135) & (1.000) & (0.922) & (0.732) & (0.780) & (0.604) \\
Stay-at-home order * Rep & -11.48*** & -14.49*** & -13.78*** & -11.97*** & -11.93*** & -12.26*** \\
 & (2.036) & (1.829) & (1.677) & (1.246) & (1.237) & (1.070) \\
School Closure & 9.670*** & 0.241 & 9.238*** & 0.189 & 0.122 & 0.274 \\
 & (0.878) & (0.426) & (0.929) & (0.337) & (0.283) & (0.289) \\
Gathering restrictions & 1.425* & -0.00126 & 2.475*** & 0.313 & 0.390* & 0.327 \\
 & (0.794) & (0.451) & (0.722) & (0.241) & (0.200) & (0.205) \\
Business closure & 0.0256 & 0.141 & 1.064** & 0.457 & 0.239 & 0.389 \\
 & (0.536) & (0.473) & (0.498) & (0.373) & (0.409) & (0.352) \\
Precipitation & 0.00739*** & 0.00399*** & 0.00731*** & 0.00449*** & 0.00267*** & 0.00372*** \\
 & (0.00104) & (0.000972) & (0.000886) & (0.000534) & (0.000726) & (0.000582) \\
Temperature & -0.0143*** & -0.00567** & -0.0203*** & -0.00999*** & -0.0116*** & -0.0125*** \\
 & (0.00235) & (0.00227) & (0.00225) & (0.000902) & (0.00148) & (0.00120) \\
Constant & 1.653*** & -0.304 & 3.343*** & -0.324 & 4.843*** & 4.917*** \\
 & (0.339) & (0.202) & (0.482) & (0.304) & (0.731) & (0.525) \\
 &  &  &  &  &  &  \\
Observations & 5,194 & 5,194 & 5,194 & 5,194 & 5,194 & 5,194 \\
R-squared & 0.723 & 0.943 & 0.753 & 0.964 & 0.979 & 0.973 \\
\hline
Time fixed-effects            & NO          & YES         & NO          & YES         & NO          & NO          \\
State fixed effects           & NO          & NO          & YES         & YES         & YES         & YES         \\
Region*time fixed effects     & NO          & NO          & NO          & NO          & YES         & NO          \\
Division*time fixed effects   & NO          & NO          & NO          & NO          & NO          & YES         \\              
\hline
\end{tabular}
\begin{minipage}[!hb]{1\linewidth}
\vspace{0.5cm}
Note: All models are OLS regressions and uses the time spent at home from Google Mobility. State-level clustered robust standard errors in parentheses with *** p$<$0.01, ** p$<$0.05, * p$<$0.1. 
\end{minipage}
\label{Table4}
\end{sidewaystable}

\section{Conclusion}
	\label{conclusion}

The results of the paper show evidence that beliefs shared on Twitter are correlated with mobility at the state level. Revealed beliefs related to social distancing on Twitter are positively correlated with the practice of social distancing at the state level. The effects remain significant and stable in magnitude when we control for additional factors. The results of this paper are helpful to disentangle the effects of voluntary responses based on beliefs from the effects of government decisions to implement social distancing policies. Social networks such as Twitter reveal the beliefs of individuals about social distancing and are a good indicator of their willingness to comply with containment policies. The results also show some differences in the magnitude of alternative Twitter indices considering the number of likes, tweets and replies. The evidence does not permit us to ultimately pin down the network effects of self-revealed beliefs on the beliefs and behaviors of other individuals and opens avenues for further research. \\

Our analysis has policy implications for flattening the curve. The results suggest that the impact of voluntary responses, based on beliefs and available information, on mobility is important and observed across states. Our results demonstrate the importance of accounting for human beliefs in designing containment policies, which is rarely considered in traditional SIR models. As beliefs and behaviors are related, excessive lockdown measures might not be useful when individuals' behavior is already precautionary and vice versa. Theoretically, this means that government orders to increase social distancing are not the only instrument to combat the negative externality of insufficient social distancing. \\

There are, of course, limitations to our analysis. The timing and the location of the beliefs displayed on social networks follows the spread of the infection and the anticipation of the risk of contamination. By the same token, state orders are not randomly assigned and result from comparing the economic costs and health outcomes of the measures. If Republicans are more reluctant to enforce social distancing, Republican governors might also be slower to adopt a social distancing policy. We cannot rule out that the spread of contamination and state orders have both direct and indirect effects, via beliefs shared on Twitter, on mobility. We do not relate our results to some potential outcomes in terms of COVID-19 cases and deaths or necessary mobility and unnecessary mobility, which we leave to epidemiologists.\\

\pagebreak    
\onehalfspacing

\bibliography{bib.bib}

\begin{thebibliography}{}

\bibitem[Abouk and Heydari, 2020]{abouk}
Abouk, R. and Heydari, B. (2020).
\newblock The immediate effect of covid-19 policies on social distancing
  behavior in the united states.
\newblock {\em medRxiv: 2020.04.07.20057356}.

\bibitem[Alfaro et~al., 2020]{alfaroetal}
Alfaro, L., Faia, E., Lamersdorf, N., and Saidi, F. (2020).
\newblock Social interactions in pandemics: Fear, altruism, and reciprocity.
\newblock {\em NBER working paper 27134}.

\bibitem[Allcott et~al., 2020]{allcott2020polarization}
Allcott, H., Boxell, L., Conway, J., Gentzkow, M., Thaler, M., and Yang, D.~Y.
  (2020).
\newblock Polarization and public health: Partisan differences in social
  distancing during the coronavirus pandemic.
\newblock {\em NBER Working Paper}, (w26946).

\bibitem[Altig et~al., 2020]{altigetal}
Altig, D., Baker, S.~R., Barrero, J.~M., Bloom, N., Bunn, P., Chen, S., Davis,
  S.~J., Leather, J., Meyer, B.~H., Mihaylov, E., Mizen, P., Parker, N.~B.,
  Renault, T., Smietanka, P., and Thwaites, G. (2020).
\newblock Economic uncertainty before and during the covid-19 pandemic.

\bibitem[Andersen, 2020]{andersen}
Andersen, M. (2020).
\newblock Early evidence on social distancing in response to covid-19 in the
  united states.
\newblock {\em Working paper}.

\bibitem[Baker et~al., 2020]{bakertwitter}
Baker, S.~R., Bloom, N., Davis, S.~J., and Renault, T. (2020).
\newblock Economic uncertainty measures derived from twitter.

\bibitem[Barrios et~al., 2020]{barriosetal}
Barrios, J.~M., Benmelech, E., Hochberg, Y.~V., Sapienza, P., and Zingales, L.
  (2020).
\newblock Civic capital and social distancing during the covid-18 pandemic.
\newblock {\em NBER working paper 27393}.

\bibitem[Brodeur et~al., 2020]{brodeur2020covid}
Brodeur, A., Clark, A., Fleche, S., and Powdthavee, N. (2020).
\newblock Covid-19, lockdowns and well-being: Evidence from google trends.

\bibitem[Bursztyn et~al., 2020]{bursztynetal}
Bursztyn, L., Rao, A., Roth, C.~P., and Yanagizawa-Drott, D.~H. (2020).
\newblock Misinformation during a pandemic.
\newblock {\em NBER working paper 27417}.

\bibitem[Carneiro and Mylonakis, 2009]{carneiromylonakis}
Carneiro, H. and Mylonakis, E. (2009).
\newblock Google trends: A web-based tool for real-time surveillance of disease
  outbreaks.
\newblock {\em Clinical Infectious Diseases}.

\bibitem[Chernozhukov et~al., 2020]{chernozhukov2020}
Chernozhukov, V., Kasahara, H., and Schrimpf, P. (2020).
\newblock Causal impact of masks, policies, behavior on early covid-19 pandemic
  in the u.s.
\newblock {\em arXiv preprint arXiv::2005.14168v2}.

\bibitem[Cronin and Evans, 2020]{cronin2020private}
Cronin, C.~J. and Evans, W.~N. (2020).
\newblock Private precaution and public restrictions: What drives social
  distancing and industry foot traffic in the covid-19 era?
\newblock Technical report, National Bureau of Economic Research.

\bibitem[Ding et~al., 2020]{dingetal}
Ding, W., Levine, R., Lin, C., and Xie, W. (2020).
\newblock Social distancing and social capital: Why u.s. counties respond
  differently to covid-19.
\newblock {\em NBER working paper 27393}.

\bibitem[Eady et~al., 2019]{eady2019many}
Eady, G., Nagler, J., Guess, A., Zilinsky, J., and Tucker, J.~A. (2019).
\newblock How many people live in political bubbles on social media? evidence
  from linked survey and twitter data.
\newblock {\em Sage Open}, 9(1):2158244019832705.

\bibitem[Ferguson et~al., 2020]{ferguson2020}
Ferguson, N., Laydon, D., Nedjati-Gilani, G., Imai, N., Ainslie, K., Baguelin,
  M., Bhatia, S., Boonyasiri, A., Cucunuba, Z., Cuomo-Dannenburg, G., Dighe,
  A., Dorigatti, I., Fu, H., Gaythorpe, K., Green, W., Hamlet, A., Hinsley, W.,
  Okell, L.~C., van Elsland, S., Thompson, H., Verity, R., Volz, E.,
  YuanrongWang, H., Walker, P.~G., Walters, C., Winskill, P., Whittaker, C.,
  Donnelly, C.~A., Riley, S., and Ghani, A.~C. (2020).
\newblock Report 9: Impact of non-pharmaceutical interventions (npis) to reduce
  covid-19 mortality and healthcare demand.
\newblock {\em Technical report}.

\bibitem[Fullman et~al., 2020]{fullman2020state}
Fullman, N., Bang-Jensen, B., Amano, K., Adolph, C., and Wilkerson, J. (2020).
\newblock State-level social distancing policies in response to covid-19 in the
  us [data file].
\newblock {\em Version}, 1:2020.

\bibitem[Gupta et~al., 2020]{gupta2020}
Gupta, S., Nguyen, T.~D., Lozano~Rojas, F., Raman, S., Lee, B., Bento, a.,
  Simon, K.~I., and Wing, C. (2020).
\newblock Tracking public and private responses to the covid-19 epidemic:
  Evidence from state and local government actions.
\newblock {\em NBER working papier 27027}.

\bibitem[Kapoor et~al., 2020]{kapoor2020god}
Kapoor, R., Rho, H., Sangha, K., Sharma, B., Shenoy, A., and Xu, G. (2020).
\newblock God is in the rain: The impact of rainfall-induced early social
  distancing on covid-19 outbreaks.
\newblock {\em Available at SSRN 3605549}.

\bibitem[Malik et~al., 2015]{malik2015population}
Malik, M.~M., Lamba, H., Nakos, C., and Pfeffer, J. (2015).
\newblock Population bias in geotagged tweets.
\newblock In {\em Ninth international AAAI conference on web and social media}.

\bibitem[Moore et~al., 2019]{mooreetal}
Moore, F.~C., Obradovich, N., Lehner, F., and Baylis, P. (2019).
\newblock Rapidly declining remarkability of temperature anomalies may obscure
  public perception of climate change.
\newblock {\em Proceedings of the National Academy of Sciences of the USA}.

\bibitem[Pavalanathan and Eisenstein, 2015]{pavalanathan2015confounds}
Pavalanathan, U. and Eisenstein, J. (2015).
\newblock Confounds and consequences in geotagged twitter data.
\newblock {\em arXiv preprint arXiv:1506.02275}.

\bibitem[Renault, 2017]{renault2017}
Renault, T. (2017).
\newblock Intraday online investor sentiment and return patterns in the us
  stock market.
\newblock {\em Journal of Banking and Finance}.

\bibitem[Simonov et~al., 2020]{simonovetal}
Simonov, A., Sacher, S.~K., Dub\', J.-P.~H., and Biswas, S. (2020).
\newblock The persuasive effect of fox news: Non-compliance with social
  distancing during the covid-19 pandemic.
\newblock {\em NBER working paper 27237}.

\bibitem[Takhteyev et~al., 2012]{takhteyev2012geography}
Takhteyev, Y., Gruzd, A., and Wellman, B. (2012).
\newblock Geography of twitter networks.
\newblock {\em Social networks}, 34(1):73--81.

\bibitem[Yilmazkuday, 2020]{yilmazkuday2020stay}
Yilmazkuday, H. (2020).
\newblock Stay-at-home works to fight against covid-19: International evidence
  from google mobility data.
\newblock {\em Available at SSRN 3571708}.

\end{thebibliography}
\bibliographystyle{apalike}
\newpage

\newpage

\section*{Appendix A}

\newpage

\section*{Appendix B}
\begin{sidewaystable}[!h]
\centering
\caption{[Appendix B] Impact of Twitter indices of social distancing weighted by likes, retweets and replies on mobility}
\vspace{0.5cm}
\begin{tabular}{L{5cm}C{2.5cm}C{2.5cm}C{2.5cm}C{2.5cm}C{2.5cm}C{2.5cm}}
 \hline
 & (1) & (2) & (3) & (4) & (5)  \\
VARIABLES & Workplaces  & Groceries and pharmacies & Retail and recreation  & Transit stations  & Parks \\ \hline
 &  &  &  &  &   \\
 SD Tweets  & -1.220*** & -0.347* & -0.471** & -1.546*** & -1.684*** \\
 & (0.169) & (0.190) & (0.225) & (0.297) & (0.558) \\
 R-squared & 0.981 & 0.920 & 0.967 & 0.945 & 0.824 \\ \hline
  &  &  &  &  &   \\
 SD Tweets * likes & -0.841*** &   -0.185 &  -0.321** &  -1.094***   & -1.285*** \\
 & (0.103) &   (0.174) &  (0.152) &  (0.214) & (0.478)    \\
 R-squared & 0.981& 0.920 & 0.967& 0.945 & 0.824 \\ \hline
 &  &  &  &  &   \\
 SD Tweets * retweets  &  -0.767*** &   -0.187 &  -0.328** &  -1.053***   & -1.404***  \\
 &   (0.0867) &   (0.168) &  (0.136) &  (0.199) & (0.405) \\
 R-squared  & 0.980& 0.920& 0.967& 0.945 & 0.824 \\ \hline
 &  &  &  &  &   \\
 SD Tweets * replies  &  -0.753*** & -0.209   & -0.369***  & -1.068***   & -1.382*** \\
 &   (0.0764) &  (0.152) & (0.136) &   (0.190) &   (0.391) \\
 R-squared  &  0.980 & 0.920& 0.967& 0.945 & 0.824 \\
  \hline
 &  &  &  &  &   \\
Observations & 5,194 & 5,194 & 5,194 & 5,194 & 5,183 \\
State FE & Yes & Yes & Yes & Yes & Yes  \\
 Division*Time FE & Yes & Yes & Yes & Yes & Yes  \\ \hline
\end{tabular}
\begin{minipage}[!hb]{1\linewidth}
\vspace{0.5cm}
Note: All models are OLS regressions with state and division*time FE. Twitter indices are standardized. State-level clustered robust standard errors in parentheses with *** p$<$0.01, ** p$<$0.05, * p$<$0.1.
\end{minipage}
\label{AppendixB}
\end{sidewaystable}
\end{document}